\documentclass[aps,prl,twocolumn,floats]{revtex4-2}
\usepackage[margin=0.8in,footskip=0.25in,paperwidth=8in,paperheight=13.5in]{geometry}
\usepackage{latexsym}
\usepackage{balance}
\usepackage{amsmath}

\usepackage{amssymb}
\usepackage{bm}
\usepackage{wasysym}
\usepackage[dvips]{color}
\usepackage{graphicx}
\usepackage{subfigure}
\usepackage{epsfig}
\usepackage{balance}
\usepackage{subfigure}
\usepackage{ulem}
\usepackage[LGR,T1]{fontenc}
\usepackage{bm}
\newcommand{\hide}[1]{}
\newcommand{\veps}{\varepsilon}
\newcommand{\crq}{ {\bf Q}_1}
\newcommand{\cbq}{{\bf Q}_2}

\def\veps{\varepsilon}

\usepackage{tikz}
\usetikzlibrary{matrix}
\usetikzlibrary{arrows}

\graphicspath{{./figures/}}
\usepackage{float}
\usepackage[colorlinks,bookmarks=true,citecolor=blue,linkcolor=blue,urlcolor=blue, breaklinks=true]{hyperref}
\usepackage[sort&compress]{natbib}
\begin{document}
\title{Graphene bilayer and trilayer Moir\'e lattice with Rashba spin-orbit coupling}
\author{Y. Avishai$^{1,2}$ and Y. B. Band$^{1,3}$ \\
%{\color {red} Proprietary material, Intended solely for Prof. H. Beidenkopf, WIS}
}
\affiliation{$^1$Department of Physics,
  Ben-Gurion University of the Negev,
  Beer-Sheva, Israel.\\
%  New York University and the NYU-ECNU Institute
%  of Physics at NYU Shanghai, 3663 Zhongshan Road North,
%  Shanghai, 200062, China, \\
  $^2$ Yukawa Institute for Theoretical Physics, Kyoto, Japan.\\
  $^3$ Department of Chemistry,
  Department of Electro-Optics, and
  The Ilse Katz Center for Nano-Science,
  Ben-Gurion University of the Negev, Beer Sheva, Israel.
}

\begin{abstract}
We study the behavior of equilibrium spin currents near the magic angles of 
twisted bilayer and trilayer graphene in the presence of Rashba spin-orbit coupling. 
There is a substantial difference in the properties of local observables in twisted graphene layers, as compared with those in single and/or un-twisted layers graphene.  Remarkably, when plotted as a function of the twist angle $\theta$, the electronic charge density and the equilibrium spin currents are non-analytic at angles that are close within 1\% to the magic angles. 
In addition to the occurrence  of a rich spin texture patterns, these findings enable the determination of magic angles within an accuracy of less than 0.01$^\circ$ in terms of an STM measurement of the local density and spin resolved measuring devices for measuring equilibrium spin currents.  
\end{abstract}

\maketitle

{\it Introduction}: 
Van Hove singularities in  twisted bilayer graphene (TBG) and  the emergence of  flat bands at certain twisting angles were first reported in Refs.~\cite{Lopes,Eva, Moon}.  A continuous model for exploring electronic structure of TBG that forms a Moir\'e lattice was developed in Ref.~\cite{Rafi} and exposed the occurrence of magic angles, $\theta_{2m}$, i.e., twist angles at which the lowest (positive) energy band is flat versus crystal momentum. Their origin was clarified in Ref.~\cite{Vishwanath,Cao,Sun}, while the symmetries and topological content of this system was analyzed in Refs.~\cite{Song1, Bernevig1, Bernevig2}. 
Recent reports have shown that this system can host correlated insulating states \cite{Cao_18, Cao_16}, unconventional superconductivity \cite{Cao_18b}, distinct Landau level degeneracies \cite{Lu_19}, emergent ferromagnetism with anomalous Hall effect and quantized anomalous Hall behavior \cite{Sharpe, Pixley, Serlin}, chirality \cite{Stauber}, valley spirals \cite{Wolf} and opto-spintronics \cite{Sierra}. The flatness of the band is very sensitive to the value of the magic angle, hence, an accurate determination  the magic angle is crucial \cite{Zeldov}. 

In the present work we consider TBG and twisted trilayer graphene (TTG) subject to a uniform perpendicular electric field that causes Rashba spin orbit coupling (RSOC) \cite{Rashba}, and substatiate the relevance of magic angles to the pertinent spin physics.  The Bloch functions $\{ \Psi_{{\bf k}}({\bf r}) \}$ of the lowest conduction band at crystal momentum ${\bf k}$
% =(k_x,k_y)=(k_x,0)$ [where $k_x \in {\cal K}$  that is the segment joining the $\Gamma$ and $M$ points of the Moir\'e lattice $M_Q$], 
are calculated and employed to determine the charge density $\rho_{\bf k} ({\bf r}) = \vert \Psi_{\bf k}({\bf r}) \vert ^2$ and equilibrium spin currents (ESC) $J_{ij;{\bf k}}({\bf r})$ as a function of the twist angle $\theta$ (here $i=x,y,z$ is the polarization direction and $j=x,y$ is the velocity direction). %These quantities are calculated on the segment ${\bf k}%=(k_x,0)$ joining the $\Gamma$ and $M$ points of the %reciprocal Moir\'e lattice. 
The main results of this work are: 
(1) On varying the twist angle $\theta$, (for fixed ${k_x}$) $\rho_{k_x} ({\bf r})$ and $J_{ij; { k_x}}({\bf r})$ are shown to have discontinuous derivatives with respect to $\theta$ at angles $\{\theta_2(k_x),  \theta_3(k_x)\}$ that are close within $0.01^\circ$ to the respective magic angles $\theta_{2m}$,and $\theta_{3m}$ (obtained by minimization of the band width). 
% Specifically, for our (realistic) set of parameters, it is found that
%1) The charge density and the ESC are non-analytic at $\theta_{2,3}(k_x)$.
% [Fig.~\ref{Fig1}(a), Fig.~\ref{Fig5}(a)]. 
Hence, measuring local density (by STM) or ESC (by opto-spintronic devices \cite{Sierra}) can serve as an excellent tool for determining the magic angles.  
(2) The symmetry $J_{yx}=-J_{xy}$ and the equalities $J_{xx}=J_{yy}=0$, valid in single layer graphene \cite{Zhang} are broken. 
(3) The relation $\theta_{3m} \approx \sqrt{2} \theta_{2m}$ is extended beyond the chiral limit. 
(4) Unlike in single or double {\it un-twisted} layer graphene, the ESC depend on the position ${\bf r}$, implying the possible occurrence of spin torque \cite{Niu,AB}.

{\it Formalism}: Here we develop the formalism for TBG, (extension for TTG is straight-forward). Consider massless 2D Dirac electrons in TBG lying in the $x$-$y$ plane with twist angles $\pm \theta/2$ subject to a uniform electric field ${\bf E}=E_0 \hat{\bf z}$.  We start from the continuous Moir\'e band model \cite{Song1} wherein there is no valley mixing. The Dirac 
${\bf K}$ points in adjacent layers 1 and 2 (denoted $\{ {\bf K}_1,{\bf K}_2\})$ are offset by the twisting angle $\theta$ \cite{Song11}. This procedure defines the Moir\'e ${\bf Q}$ lattice $M_Q$ shown in Fig.~6(b) of Ref.~\cite{Song1}), wherein the {\it red and blue points}, $\{{\bf Q}_1\}$ and $\{ {\bf Q}_2\}$ denote the ${\bf K}$ points in layers 1,2 respectively. Occasionally, ${\bf Q}$ will denote both. Adjacent ${\bf K}$ points of different layers are connected by three vectors $\{ {\bf q}_j \}$ (see Eq.~\ref{H} below). The $\theta$ dependent length of the vectors $\vert {\bf q}_j \vert =K_D=2 K \sin(\theta/2)$ is the $M_Q$ lattice constant (here $K=\vert{\bf K}_1\vert=\vert{\bf K}_2\vert$ \cite{Song12}).  The $\Gamma$ point marks the center of the unit cell, and the electron wave number is ${\bf k}=(k_x,k_y) \in $\ BZ of $M_Q$.  Practically, the number $N_{\bf Q}$ of ${\bf Q}$ points is cutoff within a circle centered at the $\Gamma$ point, thereby conserving the rotation symmetries specified in Ref.~\cite{Bernevig1}. Explicitly, $N_{{\bf Q}_1}=N_{{\bf Q}_2}=50\Rightarrow N_{\bf Q}=100$.  We denote by ${\bm \tau}$ the isospin encoding the two-lattice structure of single layer graphene, by ${\bm \eta}$ the pseudo-spin operator for the two layers and by ${\bm \sigma}$  the operator for the electron real spin. The pertinent 8 dimensional Hilbert space is then ${\bm \eta} \otimes {\bm \sigma} \otimes{\bm \tau}$.  RSOC is introduced as an SU(2) vector potential, ${\bf A}= [{\bm \sigma} \times \hat{\bf z}]$.  In ${\bf r}$ space the Hamiltonian $H=H_0({\bf r})+H_1$ is
\begin{eqnarray} \label{Hr}
H_0({\bf r})&=&\eta_0 \otimes \left [-i \sigma_0 \partial_{\bf r} +\lambda {\bf A} \right ] \cdot {\bm \tau} \nonumber \\
&-& \tfrac{\theta}{2} \eta_z \left [(-i)  \sigma_0 \partial_{\bf r}  +\lambda {\bf A} \right ]\times {\bm \tau}, \nonumber \\
H_1 &=& \eta^- \sigma_0 T^\dagger +\eta^+ \sigma_0 T.
\end{eqnarray}
$H$ is the extension of the Hamiltonian introduced in Eq.~(1) of Ref.~\cite{Song1}, with RSOC included. Here $\lambda$, which is proportional to $E_0$, is the RSOC strength, and $T$ is a 2$\times$2 matrix in ${\bm \tau}$ space (see below).  

%Mathematica File: Moire-h0.nb
% Analytic v_i({\bf p}) in Mathematica file RSOC-Moire-vpc.nb
We define shifted wave numbers ${\bf p}_\eta = {\bf k}-{\bf Q}_\eta$, ($\eta=1,2$ for layers 1,2).  Basis eigenfunctions of $H_0({\bf r})$ are $e^{i {\bf p}_\eta\cdot {\bf r}} v_i({\bf p}_\eta)$, where $\{v_i({\bf p}_\eta), (i=1,2,\ldots,8 \} $ are the eight dimensional eigenvectors of the 8$\times$8 matrix obtained after replacing $-i \partial_{\bf r} \to {\bf p}_\eta$ in Eq.~(\ref{Hr}).  The corresponding energies are $\epsilon_i({\bf p}_\eta)$.  Putting together these eight column eigenvectors defines an 8$\times$8 eigenvector matrix ${\bf v}({\bf p}_\eta)$.  Both $v_i({\bf p}_\eta)$ and $\epsilon_i({\bf p}_\eta)$ are expressible analytically. A Bloch eigenfunction of $H$ (an 8 dimensional vector), is expanded in plane-wave spinors $\{ e^{-i {{\bf Q}_\eta} \cdot {\bf r}} {\bf w}({\bf p}_\eta) \}$ [defined in Eq.~(\ref{Bloch1})] as,
\begin{eqnarray} \label{Bloch1}
&& \Psi_{{\bf k}}({\bf r})=\frac{e^{i {\bf k} \cdot {\bf r}}}{\sqrt{A}} \sum_{\eta=1}^{2} u_{\eta \bf k}({\bf r}) \nonumber \\
&& u_{\eta \bf k}({\bf r})=
\sum_{{{\bf Q}_\eta} \in M_Q} 
e^{-i {{\bf Q}_\eta} \cdot {\bf r}} 
\underbrace{[\sum_{i=1}^8 a_{i}({\bf p}_\eta)v_i({\bf p}_\eta)]}_{{\bf w}({\bf p}_\eta)}.
\end{eqnarray}
Here $A$ is the area of a unit cell (Moir\'e hexagon) in position space, and $\{ a_{i}({\bf p}_\eta) \}$ 
are $N$ (yet unknown) coefficients. The functions $\{ u_{\eta {\bf k}}({\bf r}) \}$ are dimensionless, and periodic on their respective triangular (Bravais) lattices in ${\bf r}$ space.  The Bloch functions $\{ \Psi_{\bf k}({\bf r}) \}$ and the coefficients $\{ a_{i}({\bf p}_\eta) \}$ should carry also a band number $n$ that is occasionally omitted for convenience.  Two notational definitions are useful:
(1) The 100 8$\times$8 matrices $\{ {\bf v}({\bf p}_\eta) \}$, are used to form an $N$$\times$$N$ block diagonal matrix 
\begin{equation} \label{V}
V \equiv \mbox{diag} \underbrace{[{\bf v}({\bf p}_\eta)]}_{8 \times 8}, \ \ 
{\bf p}_\eta={\bf k}-{\bf Q}_\eta, \ \  {\bf Q}_\eta \in  M_Q. 
\end{equation}
(2) The $N$ unknown coefficients on the RHS of Eq.~(\ref{Bloch1}) are arranged to form a vector (of $N$ components) ${\bf a}  \equiv \{ a_{i}({\bf p}_{\eta}) \})$, where $i=1, \dots, 8$.

The eigenvalue equation for the vector ${\bf a}$, employs the Bloch representation of the Hamiltonian ${\cal H}=V^\dagger HV$ in the presence of RSOC:
\begin{eqnarray} \label{VHV}
&& {\cal H}{\bf a}_n({\bf k})=\veps_n({\bf k}) {\bf a}_n({\bf k}), \nonumber \\
&& H_{{\bf Q},{\bf Q}'} = \frac{1}{A} \int e^{-i{\bf Q} \cdot{\bf r}}H({\bf r})e^{i{\bf Q}'\cdot{\bf r}} d{\bf r}. 
\end{eqnarray}
$H_{{\bf Q},{\bf Q}'} $ is  an $8 \times 8$ matrix in ${\bm \eta} \otimes {\bm \sigma}\otimes {\bm \tau}$ space,
and dim$[H]$= $N\times N, \ (N=N_{\bf Q} \times 8=800$). Explicitly, 
\begin{eqnarray} \label{H}
 && H= \tiny{\begin{bmatrix} 
H^0_{{\bf Q}_1,{\bf Q}_1} & H^1_{{\bf Q}_1,{\bf Q}_2}\\H^{1 \dagger}_{{\bf Q}_1,{\bf Q}_2}& H^0_{{\bf Q}_2,{\bf Q}_2}
\end{bmatrix}}, (\crq,\cbq)=1,2,\ldots,N_{\bf Q}, \nonumber \\
&& H_{{\bf Q},{\bf Q}'}  \equiv H^0_{{\bf Q},{\bf Q}}+H^1_{{\bf Q},{\bf Q}'} \nonumber \\
&&  H^0_{{\bf Q},{\bf Q}}({\bf k})= \eta_0 \otimes  [\sigma_0 {\bf p}+\lambda {\bf A}] \cdot {\bm \tau} \nonumber \\
&& -\tfrac{1}{2} \theta \, \zeta_{{\bf Q}} \delta_{{\bf Q},{\bf Q}'} \eta_z \otimes [\sigma_0 {\bf p}+\lambda {\bf A}] \times {\bm \tau} \nonumber \\
&&H^1_{{\bf Q},{\bf Q'}}({\bf k})=\eta^- \otimes \sigma_0 T^\dagger_{{\bf Q},{\bf Q}'}+ \eta^+ \otimes \sigma_0 T_{{\bf Q},{\bf Q}'}, \nonumber \\
&&T_{{\bf Q},{\bf Q}'}= \Sigma_{j=1}^3 [\delta_{{\bf Q}-{\bf Q}'},{{\bf q}_j}+\delta_{{\bf Q}'-{\bf Q}},{{\bf q}_j}]T_j \nonumber \\
&& {\bf q}_j = K_D\left [ \cos\tfrac{ (4j-3)\pi}{6} \hat{\bf x}+ \sin \tfrac{(4j-3)\pi}{6} \hat{\bf y} \right ], \nonumber \\
&& T_j=w_0 \tau_0+w_1 [ \cos \tfrac{2 \pi (j-1)}{3} \tau_x+  \sin \tfrac{2 \pi (j-1)}{3} \tau_y ],
\end{eqnarray} 
which is the extension of Eq.~(A3) in Ref.~\cite{Song1}.  As far as the spectrum is concerned, diagonalization of $H_{{\bf Q},{\bf Q}'}$ is sufficient. For calculating wave functions, the eigenvectors $\{ {\bf a} \}$ are required from the solution of the first of Eq.~(\ref{VHV}), to be used in Eq.~(\ref{Bloch1}). 

{\it Results for TBG}: 
%Moire-all4-incoherent_Smalltheta-new.nb 29//08/2021.
%Mathematica files in 1) Downloads, 2) YA-Backup, 3) RSOC-Moire/Mathematica-Moire-Bernevig
%MOST IMPORTANT MATEHEMATICA FILE: Moire-Direct-Bloch-100-SO-Lowest-4-levels.nb
%Moire-Direct-Bloch-100-Chiral-KD.nb
%Figures from RSOC-Moire-Figures1-Smaller-KD
%Moire-Direct-Bloch-100-Chiral-w0=0-norm-SO
%Moire-Direct-Bloch-100-Chiral-KD-norm-NSO
%Moire-Direct-Bloch-100-Chiral-KD-norm-SO
The spectrum $\{ \veps_n({\bf k}) \}$ depends on the potential parameters $w_0,w_1,\theta,\lambda$. It is calculated on the segment 
\begin{equation} \label{calK}
{\bf k} \in {\cal K}\equiv [0 \le k_x \le k_D\tfrac{\sqrt{3}}{2}, k_y=0], 
\end{equation} 
joining the $\Gamma$ and $M$ points in $M_Q$ \cite{Song13}.  The spin observables depend on ${\bf k}$ and, (unlike the case of un-twisted layers), on the position ${\bf r}=(x,y)$ in the unit cell (due to the presence of the coupling matrices $\{ T_j \}$).  We use the following parameters: 
 $K=15.0533$ (nm)$^{-1}$\cite{Vishwanath}, $w_0=77.0371$ meV, $w_1=110.053$ meV $\lambda=1.0544 \approx 1$ meV~\cite{Gmitra0,Sergej1,Gmitra}.  
%For twist angles around 1$^\circ$, $K_D = 2 k_G \sin \tfrac{\theta}{2} \approx 0.262726$ (nm)$^{-1}$.  

Our first task is to find the magic angle $\theta_{2m}$. There are different criteria for its determination, such as vanishing of the Dirac speed, minimal bandwidth, or maximal band gap to higher bands.
Ideally, the lowest band at the magic angle  is flat, but within a numerical scheme on a system of finite size the situation is less simple. For every twist angle $\theta$, the lowest conduction band 
$\veps_0(k_x,\theta)>0$ depends weakly on $k_x \in {\cal K}$. 
Then $\theta_{2m}$ may be defined as the twist angle that minimizes the difference
\begin{equation} \label{dtheta}
d(\theta) \equiv \mbox{Max}[\veps_0(k_x,\theta)]-\mbox{Min}[\veps_0(k_x,\theta)] .
\end{equation}
Using this criterion, we find the lowest magic angle to be $\theta_{2m}$=1.099$^\circ$.
%d(theta).nb
% for which $d(\theta_{2m})$=0.34 meV. 
%Moire-Spectrum-Withg-SO-near-0.nb 
%Here we expose the role of $\theta_{2m}$ in the behavior local observables such as charge density and, 
%most importantly, ESC. Explicitly, it is found that, when plotted as function of $\theta$, 
%the density $\vert \Psi_{0 {\bf k}}({\bf r}) \vert^2$ and the ESC have discontinuous derivatives at $\theta_{2m}$ [as in Fig. \ref{Fig1}(a)] for the density, and Figs.~\ref{Fig3-ab}(a) and (b) for the ESC]. 
The fact that the band is not perfectly flat, (so, strictly speaking, different Bloch functions $\{ \Psi_{ {\bf k}}({\bf r}) \}$  are {\it not degenerate}), poses the question of how to interpret the results obtained for different crystal momenta ${\bf k}$.  Recall that for single layer graphene \cite{Zhang}, where the energy depends solely on $k^2$, it is possible to integrate the ESC over a constant energy line that is a circle in the ${\bf k}$ plane. This integration procedure over a constant energy line is not applicable here, and the results should be analyzed for each ${\bf k}$ separately. Fortunately,  there are two factors that can ameliorate this obstacle: (1) There is a substantial progress in designing momentum resolved experiments  \cite{Tusche_15, Unzelmann_21,Avsar,Zollner,Wang_08}, and (2) The angles $\{\theta_2(k_x)\}$ at which the ESC displays non-analyticity are close to  $\theta_{2m}$ (within $1\%$).  
%Jxx-Jxy-kx.nb
Therefore, for definiteness, in the following we will present our results for the density and the ESC at ${\bf r}=0$, (the center of the direct Moir\'e lattice unit cell), and for fixed ${\bf k}=(k_x,k_y)=(0.05,0)$ (nm)$^{-1} \in {\cal K}$, Eq.~(\ref{calK}). This choice is convenient because, as shown below, somewhat accidentally, at this specific wave number, $\theta_2(k_x=0.05) = 1.0984^{\circ}$ just below $\theta_{2m}=1.0990^\circ$.

\begin{figure}%[htb]
%Mathematica File Moire-Spectrum-With-SO-Circle-Final.nb
%Last one Moire-Spectrum-With-SO-Final [October 16 2021];
\centering
%\subfigure[]
{\includegraphics[width=0.95\linewidth]{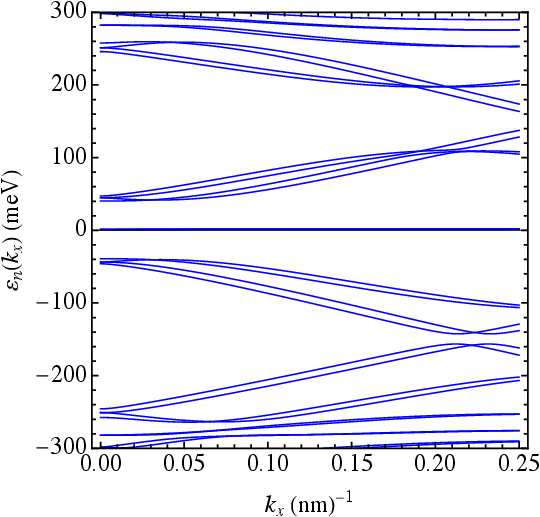}} 
\caption{\footnotesize 
 Low energy spectrum of TBG at $\theta_{2m}=1.099^\circ$ for ${\bf k} \in {\cal K}$ in the presence of RSOC. Here $\lambda \approx 10$ meV is intentionally enlarged so that the SO splitting is clearly visible.  As $\lambda \to 0$, the RSOC splitting shrinks and the level pattern is commensurate with that of Fig.~1(d) in  Ref.~\cite{Song1}.}
\label{Fig1}
\end{figure}

%Mathematica File: Moire-Spectrum-With-SO-Circle-10/18/2021.nb
%Mathematica File: Moire-Spectrum-No-SO-Circle-Play1.nb 05/09/2021
%See figure in Mathematica File Moire-Spectrum-With-SO-Circle-small-degeneracy.nb
The spectrum of several levels above and below the (nearly) flat band is plotted in Fig.~\ref{Fig1}
for $\lambda \approx 10$ meV ($\lambda$ is intentionally taken to be much larger than realistic values in order to make the SO splitting visible). Compare with the spectrum for $\lambda=0$, shown in Fig. 1(d) of Ref.~\cite{Song1}. 
%{\color{red} Moire-Bloch-Functions-YA.nb RSOC-Moire-Non-%Degenerate.nb}

%\underline{Degeneracy}:
% From Mathematica File Moire-Direct-Bloch-Check-Degeneracy.nb
%For the parameters specified above, and wave number ${\bf k}=(0.05,0)$ (nm)$^{-1}$, the four lowest positive energy levels in the conduction band 
%%are [1.33006, 1.33006, 1.32825, 1.32825] meV (i.e., they 
%are composed of two pairs of degenerate levels. However, compared with the RSOC energy strength $\lambda \approx 1$ meV, the spacing between the two pairs is  very small (a few $\mu$eV). Hence, one can regard them as four degenerate levels.  This is relevant for the calculation of spin observables, (see below). 
%See figure in Mathematica File Moire-Spectrum-With-SO-Circle-small-degeneracy.nb

\begin{figure}%[htb]
%Moire-all4-incoherent_Smalltheta-new.nb 29//08/2021.
%Moire-all4-incoherent_Tiny_theta-kx=0.07
\centering
\subfigure[]
{\includegraphics[width=0.9\linewidth]{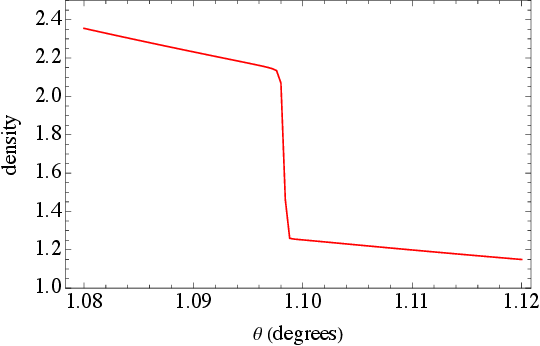}} 
\subfigure[]
{\includegraphics[width=0.9 \linewidth]{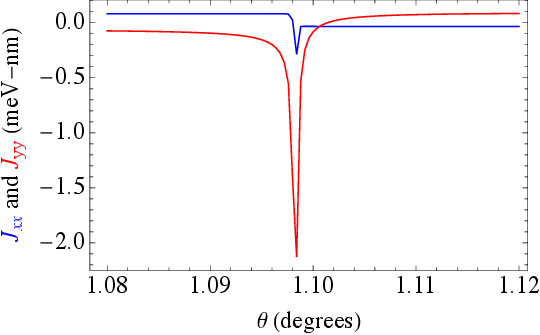}} 
%Mathematica File Moire-all4-incoherent_Smalltheta
%Mathematica FileMoire-k-ne-0-Finite-r.nb
%Mathematica File: Moire-k-ne-0-Finite-r-Direct-Bloch'.nb (the last one). 
%Mathematica File: Moire-Direct-Bloch-100-SO-10-15.nb (15/10/2021)
\centering
\subfigure[]
{\includegraphics[width=0.9\linewidth]{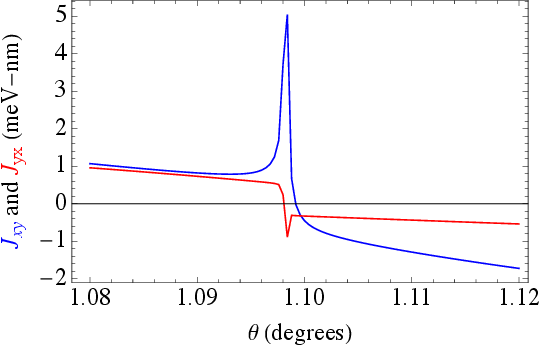}} 
\caption{\footnotesize  Local observables of TBG for $\lambda=1$ meV as function of $\theta$, displaying singularities at 
$\theta_{2}(k_x=0.05) =1.0984^{\circ}$ that is very close to the magic angle  $\theta_{2m}=1.099^\circ$.
(a) Dimensionless charge  density $A \Psi^\dagger_{{\bf k}}({\bf 0})\Psi_{{\bf k}}({\bf 0})$ showing a step  (A similar step {\it with the same} $\theta_{2}(k_x)$ occurs also for $\lambda=0$). 
(b) $J_{xx}$ and $J_{yy}$. (c) $J_{xy}$ and $J_{yx}$. }
\label{Fig2-ab}
\end{figure}

%\underline{Spin operators and spin observables}: 
%While the spectrum shown above just displays SO splitting at the magic angle, 
 Density and spin observables in TBG are {\it local}, and expressed in terms of the Bloch functions $\Psi_{{\bf k}}({\bf r})$, Eq.~(\ref{Bloch1}), and  pertinent operators  $\hat {o}$ that are 8$\times$8 matrices in ${\bm \eta} \otimes {\bm \sigma} \otimes {\bm \tau}$ space.  The spin and velocity operators are,
\begin{equation} \label{spinop}
\hat{\bf s}= \tfrac{1}{2}\hbar \eta_0 \otimes {\bm \sigma} \otimes  \tau_0, \ \hat {\bf v}=\eta_0 \otimes \sigma_0 \otimes {\bm \tau}.
\end{equation}
%the Dirac velocity operator 
%\begin{equation} \label{vop}
% \hat {\bf v}=\eta_0 \otimes \sigma_0 \otimes {\bm \tau},
%\end{equation}
The ESC tensor operator is,
\begin{eqnarray} \label{Jtensor}
%&& \mathbb {J}_{ij}= \tfrac{1}{2} [\hat {S}_i \hat {V}_j+\hat {V}_j\hat {S}_i], \ (i=x,y,z, \ \ j=x,y). 
%\nonumber \\
&& \mathbb {J}_{ij}= \tfrac{1}{2} [\hat {s}_i \hat {v}_j+\hat {v}_j\hat {s}_i], \ (i=x,y,z, \ \ j=x,y).
\end{eqnarray}
In case of $m$-fold degeneracy (for fixed ${\bf k}$), the $m$ degenerate eigenfunctions contribute {\it incoherently} to the pertinent observable, %With the parameters specified above we check that at $\theta(k_x=0.05)$, the four lowest positive energy %levels 
% in the conduction band are composed of two pairs of degenerate levels. 
%%Mathematica file: Moire-Direct-Bloch-100-degeneracy-at-tm.nb
%However, compared with the RSOC energy strength $\lambda \approx 1$ meV, the spacing between the two pairs is  very small (a few $\mu$eV). Hence, one can regard them as four 
%are nearly degenerate, i.e., 
%$m=4$. 
\begin{equation} \label{Observables}
O_{\bf k}({\bf r})=\frac{1}{m} \sum_{n=1}^m\Psi^\dagger_{{\bf k}n}({\bf r}) \hat {o} \Psi_{{\bf k}n}({\bf r}). 
\end{equation}
For the charge density, $\hat{o}={\bf 1}_{8 \times 8}$. Figure~\ref{Fig2-ab}(a) shows the non-analyticity at $\theta_2(k_x=0.05)=1.0948^\circ$ just below $\theta_{2m}=1.099^\circ$. For the spin polarization ${\bf S}$, $\hat{o}=\hat{s}$, but due to (non-trivial) time reversal invariance the {\it measured polarization should vanish}.  
%Note that when the spin operator $\hat{\bf s}$ is inserted into Eq.~(\ref{Observables}), the result does not vanish.  This is because
 The model of Ref.~\cite{Rafi} is uniquely specified for the ${\bf K}$ valleys of the two layers from which $M_Q$ is constructed. But time reversal maps ${\bf K} \to {\bf K}'$ so that each Bloch function $\Psi_{n{\bf k}}({\bf r})$ built for the Moir\'e lattice of points ${\bf K}$ in Eq.~(\ref{Bloch1}) has its Kramers partner $\Psi'_{n{\bf k}}({\bf r})$ built for the Moir\'e lattice of points ${\bf K}'$. Due to time reversal invariance the sum of the contributions of the two functions to the spin polarization vanishes.
%As shown in Fig.~\ref{Fig2}, the planar polarizations $S_x$ and $S_y$ ha{\bf K}'$.   very sharp kinks at $\theta_{2m}=1.0984^{\circ}$ (this angle was used in evaluating the spectrum in Figure~\ref{Fig1}). It is remarkable (and experimentally important) to note that for $\lambda \approx 1$ meV, the polarization reaches about $0.50 \hbar$.  Away from $\theta_{2m}$, $S_x$ and $S_y$ are almost constant with $\theta$.  (The perpendicular polarization $S_z$ vanishes identically). 
%
%\begin{figure}[htb]
%%Mathematica file: Moire-k-ne-0-Finite-r-Direct-Bloch'.nb
%\centering
%{\includegraphics[width=0.9\linewidth]{Fig2.eps}} 
%\caption{\footnotesize  
%Results for TBG: Planar polarizations $S_{x}$ and $S_{y}$ [defined in Eqs.~(\ref{spinop}) and Eq.~(\ref{Observables})], in the lowest positive energy state for $(k_x,k_y)=(0.05 \, \text{nm}^{-1}, 0)$ (i.e., to the right of the $\Gamma$ point) versus twist angle $\theta$.  The sharp kinks (of width $\approx 10^{-4}$ degrees) determine the magic angle $\theta_{2m}=1.0984^{\circ}$, which is then used to compute the spectrum plotted in Fig.~\ref{Fig1}(b).  
%}
%\label{Fig2}
%\end{figure}
%%Mathematica  file: Moire-all4-Smalltheta-Matrices-S.nb
%\underline{Equilibrium spin current}: 
In contrast, the ESC is even under time reversal and hence it can be calculated within the present model wherein the Moir\'e lattice is built solely from the ${\bf K}$ points of the two layers. 
%The spin current (tensor) operator is then,
%\begin{eqnarray} \label{Jtensor}
%%&& \mathbb {J}_{ij}= \tfrac{1}{2} [\hat {S}_i \hat {V}_j+\hat {V}_j\hat {S}_i], \ (i=x,y,z, \ \ j=x,y). 
%%\nonumber \\
%&& \mathbb {J}_{ij}= \tfrac{1}{2} [\hat {s}_i \hat {v}_j+\hat {v}_j\hat {s}_i], \ (i=x,y,z, \ \ j=x,y),
%\end{eqnarray}
%which needs to be inserted into Eq.~(\ref{Observables}) in order to calculate the observed ESC, $J_{ij}$.  
We find that the  perpendicular components vanish, $J_{zx}=J_{zy}=0$, but the planar components are finite. 
%Mathematica  file: Moire-all4-Smalltheta-Matrices-S.nb
The diagonal planar components $J_{xx}, J_{yy}$,  are plotted in Fig.~\ref{Fig2-ab}(b), while the non-diagonal planar components, $J_{xy}$ and $J_{yx}$  are plotted in Fig.~\ref{Fig2-ab}(c). The singularitries occur at the angle $\theta_2(k_x=0.05)=1.0984^\circ$.   In single layer graphene \cite{Zhang}, $J_{xx}=J_{yy}=0$ and $J_{xy}=-J_{yx}$.  Here, these symmetries are broken.    

%%%%%%%%%%%%Chiral Limits%%%%%%%%%%%%%%%%
{\it First Chiral limit}:  References~\cite{Vishwanath, Song1, Bernevig1, Bernevig2} showed that in the continuous model of TBG there is an approximate anti-unitary particle-hole symmetry operator ${\cal P}$ that becomes exact in the first chiral limit, $w_0 \to 0$. 
%In the presence of RSOC, in this limit, $\veps_n({\bf k},\lambda)=\veps_n({\bf k},-\lambda)$ and $\veps_n({\bf k},\lambda) \approx -\veps_n(-{\bf k},\lambda)$. 
%%%Moire-Spectrum-With-SO-Circle-Final-Chiral-KD
%%%Mathematica File: Moire-Spectrum-With-SO-Chiral-phsymmetry.nb
%it is anticipated that this operator satisfies, ${\cal P}H({\bf k},\lambda){\cal P}^{-1} \approx -H(-{\bf k},-\lambda)$, where the matrix elements of $H({\bf k},\lambda)$ are given in Eq. (\ref{H}).  
In this limit, the  ESC vanish (together with the spin polarization), and there are no relevant spin observables. By minimizing the lowest positive band width it is found that 
$\theta_{2m- \text{chiral}}=1.0887^\circ<\theta_{2m}=1.099^\circ$. 
%The spectrum is shown in Fig.~\ref{Fig3-ab}(a). Note the (almost) exact particle-hole symmetry 
%(the particles and hole bands are virtually symmetric around $\veps=0$).
The density 
(not shown in here) 
%Fig.~\ref{Fig3-ab}(b) 
is non-analytic at the (somewhat smaller) angle, namely, 
$\theta_{2- \text{chiral}}(k_x=0.05)=1.0845^\circ <\theta_{2m-\text{chiral}}=1.0887^\circ.$ 
Thus, the magic angles depend weakly on $w_0$ (for $w_0=77$ meV$, \theta_{2m}=1.099^\circ$ and for $w_0=0$ meV$, \theta_{2m-\text{chiral}}=1.0887^\circ$).  

\begin{figure}%[htb]
%Tri-Moire-10-1-21-no-chiral.nb
%Tri-Moire-10-1-21.nb
%Results-Tri-Moire-10-1-21.nb
\centering
\subfigure[]
%Tri-Moire-10-1-21.nb
%Results-Tri-Moire-10-1-21.nb
{\includegraphics[width=0.8 \linewidth]{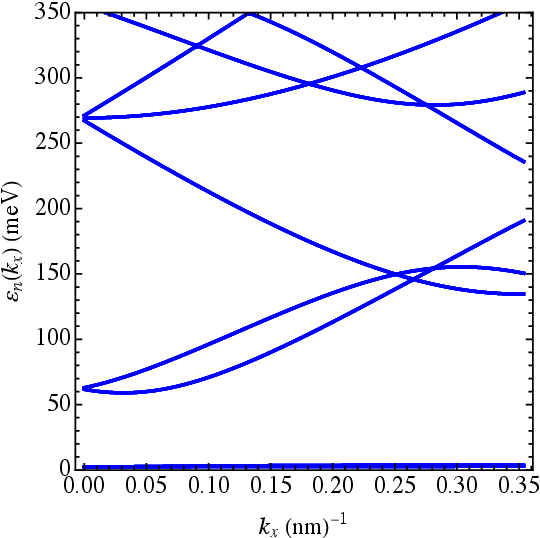}} 
\centering
\subfigure[]
%Tri-Moire-Direct-Bloch-spectrum.nb
%Results-Tri-Moire-10-1-21.nb
{\includegraphics[width=0.8\linewidth]{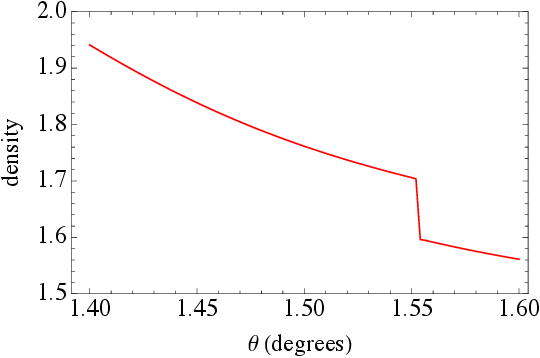}} 
\centering
\subfigure[]
%Tri-Moire-Direct-Bloch-spectrum.nb
%Results-Tri-Moire-10-1-21.nb
{\includegraphics[width=0.8\linewidth]{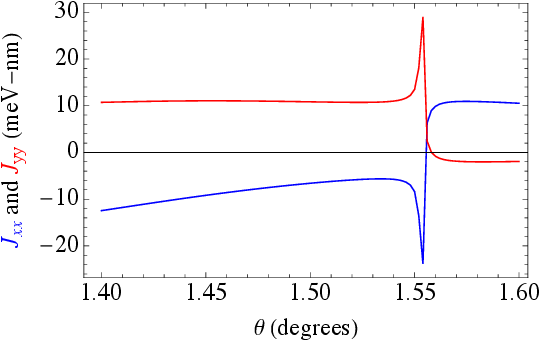}} 
\centering
\subfigure[]
%Tri-Moire-Direct-Bloch-spectrum-nochiral.nb
%Results-Tri-Moire-10-1-21.nb
{\includegraphics[width=0.8\linewidth]{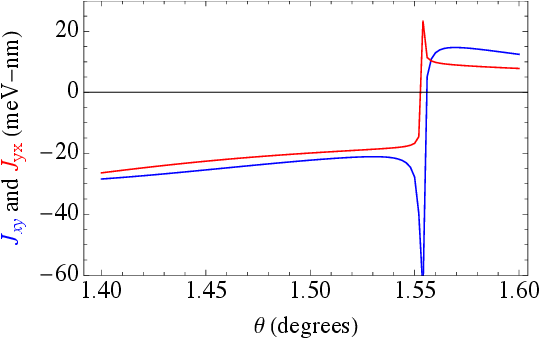}} 
\caption{\footnotesize  Results for TTG: Parameters are as in Fig.~(\ref{Fig2-ab}). 
(a) The spectrum at $\lambda=0$ for ${\bf k}\in {\cal K}$ and $\theta_{3m}=1.5545^\circ$ displays a narrow band close to $\veps=0$, separated by a large gap of about 60 meV from the next band.  For $\lambda=1$ meV there is a small SO splitting that is invisible on this scale.
 (b) Dimensionless charge density $A \Psi^\dagger_{{\bf k}}({\bf 0})\Psi_{{\bf k}}({\bf 0})$ versus $\theta$.   This pattern is non-analytic at $\theta_3(k_x=0.05)=1.5536^\circ$.  Note that $\theta_3(k_x=0.05) \approx \sqrt{2} \, \theta_2(k_x=0.05)$.  Thus, the relation $\theta_{3m} \approx \sqrt{2} \theta_{2m}$ 
 (derived in Ref.~\cite{Vishwanath} {\it in the chiral limit}) is extended to the  case $w_0>0$.  
(c) and (d) $J_{xx},J_{yy},J_{xy},J_{yx}$ displaying non-analyticity at $\theta_3(k_x=0.05) \approx \theta_{3m}$. 
}
\label{Fig4}
\end{figure}

{\it Twisted three-layer graphene with RSOC}:
%Tri-Moire-Direct-Bloch-spectrum.nb
%Tri-Moire-10-1-21.nb
%Results-Tri-Moire-10-1-21.nb
Recently, interest has grown in twisted multilayer graphene \cite{Park, Hao, Cao, Simon}. We shall now briefly address the ESC pattern and the energy spectrum in TTG. As in Ref.~\cite{Hao}, we consider a model of alternating-twist three-layer graphene for which the relative twists between two neighboring layers have the same magnitude but alternate in sign (see Fig.~1 therein).  Like in the case of TBG, we show that as function of $\theta$, the density and ESC are non-analytic at the three layer angle $\theta_3(k_x=0.05)$. We also extend a remarkable relation suggested (within the chiral limit) in Ref.~\cite{Hao} relating $\theta_{2m}$ and $\theta_{3m}$. Calculation of the spectrum and ESC are carried out for the same parameters as for the case of TBG. However, for numerical expediency, we slightly decrease the cutoff used for TBG to include 84 (instead of 100) ${\bf Q}$ points, so the Hamiltonian matrix is $N$$\times$$N$ with $N=1008$.  
 Using the notation in Eq.~(\ref{H}), the Hamiltonians of the TTG system is compactly written as,
\begin{eqnarray}   \label{H3}
H(3) = \begin{bmatrix} 
H^0_{\crq,\crq} & H^1_{\crq,\cbq} & 0 \\
 H^{1\dagger}_{\crq,\cbq} & H^0_{\cbq,\cbq}&H^1_{\cbq,\crq} \\
0&H^{1\dagger}_{\cbq,\crq} &H^0_{\crq,\crq}
\end{bmatrix}, 
\end{eqnarray}
where $\crq$ and ,$\cbq$ run on 1,2,\ldots,42). The technique for extracting spin observables requires a simple extension of the procedure used above for TBG. The Hilbert space is now ${\bm \Sigma} \otimes {\bm \sigma}  \otimes {\bm \tau}$ (${\bm \Sigma}$ is the vector of spin 1 matrices encoding the three layers), so that each Bloch function is now a 12 component plane wave spinor (after replacing ${\bm \eta} \to {\bm \Sigma}$ in the appropriate expressions). 
By inspecting the minimum of $d(\theta)$, it is found that the TTG magic angle is $\theta_{3m}$=1.5545$^\circ$. In analogy with the TBG system, it is expected that $\theta_3(k_x=0.05)$ 
(where the local observables are singular) is very closely below $\theta_{3m}$.  
This is indeed the case:
The charge density for the TTG is shown in Fig.~ \ref{Fig4}(a), while the planar components of the ESC are shown in Figs.~ \ref{Fig4}(b,c).  All the three observables display  a singularity at the angle 
$\theta_3(kx=0.05)=1.5536^\circ$. 
Therefore, 
the magic angles of the two and three layers systems are in excellent accord with the relation derived in Ref.~\cite{Hao} {\it in the first chiral limit}, namely, the relation $\theta_{3m}\approx \sqrt{2} \theta_{2m}$ is extended to the case $w_0>0$. 
 Finally, the spectrum of the TTG system at the magic angle $\theta_{3m}$ is shown in Fig.~\ref{Fig4}(d). It is characterized by a narrow band just above $\veps=0$ followed by a gap of about 60 meV.

{\it Summary}: 
In this work we considered TBG and TTG subject to RSOC. For TBG, using the criterion of minimal bandwidth we determined the magic angle and the spectrum [see Fig.~\ref{Fig1}], and then analyzed the behavior of charge density and ESC as function of the twist angle $\theta$. The fact that the band is not ideally flat requires separate analysis for each crystal momentum $k_x$.  It is shown for $k_x=0.05$/nm, but we checked that for any fixed $k_x \in {\cal K}$,  the charge density and the ESC are non-analytic as a function of the twist angle $\theta$ as it passes through a certain angle $\theta_2(k_x)$ that is {\it close to the magic angle within 0.01$^\circ$},  see Fig.~\ref{Fig2-ab}. 
%Explicitly, for the parameters $w_0,w_1$ used here (and elsewhere), $\lambda \approx 1$ meV, $(k_x,k_y)=(0.05,0)$ (nm)$^{-1}$, it is found that: 
%(1) The charge density has a steep step at the magic angle. 
%(2) The planar components of the ESC have sharp kinks at the magic angle.  Both properties (1) and (2) determine the magic angle $\theta_{2m}=1.0984^\circ$  with very high precision. With this angle, the criteria of narrow band and large gap are convincingly satisfied,  
Symmetry relations among ESC components displayed in single layer graphene \cite{Zhang}  and un-twisted bilayer graphene are broken in the twisted system. The reason is that in single layer graphene, the ${\bf k} \cdot {\bf p}$ expansion around the Dirac points is assumed \cite{Zhang}. The ${\bf k} \cdot {\bf p}$  model {\it has a continuous rotation symmetry, which is higher than the discrete symmetries of the TBG. This rotational symmetry is broken in TBG. 
 Unlike in Ref.~\cite{Zhang}}, all spin observables depend on the position ${\bf r}$, implying the possible occurrence of spin torque \cite{Niu,AB}. The pattern of density and ESC is displayed here for ${\bf r}=0$ but $\theta_{2}(k_x)$ is independent of ${\bf r}$ (within the unit cell).

An analogous study with similar results is shown for TTG, wherein the respective angles $\theta_3(k_x=0.05)$ and $\theta_{3m}$ are related to  $\theta_2(k_x=0.05)$ and $\theta_{2m}$ by a factor $\approx \sqrt{2}$. This extends the relation $\theta_{3m} \approx \sqrt{2} \theta_{2m}$ claimed in Ref.~\cite{Hao} in the chiral limit ($w_0=0$) also for $w_0>0$.
   
Thus, in addition to the association of magic angles with flat bands, correlated insulating states, unconventional superconductivity, ferromagnetism with anomalous Hall effect and distinct Landau level degeneracies, they are also relevant to spin physics. Following the recent developments in the research of monolayer and (un-twisted) multilayer graphene spintronics \cite{Avsar, Zollner}, we hope our study will stimulate experimental and further theoretical work on the role of magic angles to the spin physics of Moir\'e systems. This expectation is corroborated by the hope that
ESC can be measured using spin and angle-resolved photoelectron spectroscopy \cite{Tusche_15, Unzelmann_21, Avsar, Zollner} and polarized light scattering \cite{Wang_08}.

\begin{acknowledgments}
We are grateful to Rafi Bistritzer, Zhi-Da Song, Pilkyung Moon, Alex Kruchkov and Ady Stern for useful discussions.
\end{acknowledgments}

\end{document}